\documentclass[amsmath,amssymb,prc,preprint]{revtex4-2}

\usepackage{amssymb,amsmath} 
\usepackage{graphicx,epsfig} 

\newcommand{\Hh}{{}^3\mathrm{H}}
\newcommand{\He}{{}^3\mathrm{He}}
\newcommand {\mbf}[1]{{\mathbf{#1}}}

\newcommand {\mcu}{\mathcal{U}}
\newcommand {\mct}{\mathcal{T}}

\begin{document}

\title{Examination of the multiple-scattering expansion in the four-nucleon system}


\author{A. Deltuva}
\email{arnoldas.deltuva@tfai.vu.lt}
\affiliation
{Institute of Theoretical Physics and Astronomy, 
Vilnius University, Saul\.etekio al. 3, LT-10257 Vilnius, Lithuania
}

\received{January 4, 2025}

\begin{abstract}%
The elastic neutron-${}^3\mathrm{H}$ scattering at intermediate energies is studied
using rigorous integral equations solved in the momentum-space partial-wave basis.
The four-particle transition operators are expanded into multiple-scattering series
in terms of  subsystem transition operators.
Various approximations resulting from truncation of the series in different ways
are evaluated and their  validity is investigated. They fail at lower energies but
at higher energies  provide a rough reproduction of exact results at small scattering angles.
In the large-angle region all approximations fail heavily, indicating that the
scattering amplitude results from a delicate interplay of many multiple-scattering terms.
The  partial-wave analysis reveals that the developed approximations are reliable
in higher partial waves, and for practical calculations an efficient
``hybrid'' approach is proposed,
  combining exact amplitudes in lower partial waves with approximations in higher
 partial waves.
 The implications for often used  approximation of the first order in two-body transition matrix
 and development of microscopic optical potentials are discussed.
\end{abstract}

\maketitle

\section{ Introduction} 

Nucleon-nucleus scattering is a  many-body problem in the continuum.
The presence of open many-cluster channels is reflected in nontrivial boundary conditions for the
system wave function and renders the solution much more complicated than in the case of bound states.
While at very  low energies below the breakup threshold some bound-state like methods have been applied 
\cite{nollett:07a,quaglioni:08a,navratil:12a,hupin:19a}, rigorous scattering calculations
are limited so far to five-body system \cite{lazauskas:18a} but also restricted to the energy
regime below the breakup threshold. An exact treatment of the continuum in the many-channel regime 
above the breakup threshold is available presently in systems consisting of up to four nucleons
\cite{deltuva:12c,deltuva:14a,deltuva:15a,lazauskas:15a}. Thus, the four-nucleon system could be taken
as a test case to evaluate the reliability of approximate methods used for the description of many-body scattering.
At higher energies a Born-type approximation is quite common approach, relying
on the assumption that the interaction time is very short and, consequently, a single interaction only takes place.
In other words, one could expand the full transition amplitude into
multiple scattering series and retain only the lowest-order terms
\cite{CRESPO200626,PhysRevC.74.044616}. Typically, those terms
include just two-particle transition operators and can be calculated relatively easy.
The aim of this work is to estimate the accuracy of multiple scattering series
truncated in several ways
by comparing with the results of exact four-nucleon scattering calculations.
On the other hand, if a good agreement is established under particular conditions, this would enable
to replace complicated and time consuming  exact calculations by simpler ones of the Born type.

Furthermore, another very common approach to the description of the nucleon-nucleus elastic
scattering is via optical potentials.
The progress in many-nucleon structure calculations using various methods
such as the microscopic mean field, the no-core shell model, Green's function Monte Carlo, coupled cluster approach,
self-consistent Green's function has open the doors
for microscopic calculations of the optical potentials \cite{microOP:23}.
Quite a typical approach is taking nuclear densities from many-body calculations and 
folding  with the nucleon-nucleon transition matrices that act between external and internal nucleons
\cite{gennari:18a,vorabbi:22a,furumoto:mgop,PhysRevC.102.034606,PhysRevC.109.034613,PhysRevC.110.034605},
which closely resembles the lowest-order terms in multiple scattering series. Getting insight into the convergence
of the multiple scattering series could be of  use also for approaches constructing
microscopic optical potentials.

Thus, the study of the multiple scattering expansion
 could provide guidance for future calculations both in four-nucleon and heavier systems
in several ways. However, to achieve this goal an extension of existing four-nucleon scattering calculations
to  higher energies is needed. Well-converged results are available only up to about 35 MeV nucleon beam energy
  \cite{deltuva:13c,deltuva:15c} and tentative ones up to 100 MeV \cite{PhysRevC.106.054002}.
For the present study the neutron-triton scattering is chosen, where 
 the microscopic optical potential is under development \cite{vorabbi:pc}.

 Section \ref{sec.nh} recalls the four-nucleon scattering equations and develops their multiple-scattering
 expansions. Section \ref{sec:res} presents the results, while Sec.~\ref{concl} contains conclusions.
 
\section{Four-nucleon equations and their multiple-scattering expansion \label{sec.nh}}

The microscopic description of the  four-nucleon scattering in this work
is based on the Alt, Grassberger and Sandhas (AGS) equations 
\cite{grassberger:67} for transition operators $\mcu_{\beta \alpha}$. It is
 a momentum-space integral equation
 formulation of the four-particle Faddeev-Yakubovsky (FY) theory \cite{yakubovsky:67}.
 In the past several benchmark calculations for the neutron-$\Hh$, proton-$\He$,
 neutron-$\He$ and proton-$\Hh$ scattering below the breakup threshold \cite{viviani:11a,viviani:17a}
 have been performed, establishing a good agreement between  the 
 results obtained from the AGS equations,
 the hyperspherical harmonics  expansion method
\cite{viviani:01a,kievsky:08a} and the coordinate-space FY equations 
\cite{lazauskas:04a,lazauskas:09a}.
Making use of the isospin formalism, the 
symmetrized form of the AGS equations \cite{deltuva:07a}
for the nucleon-trinucleon scattering at the available energy $E$ in the center-of-mass (c.m.) frame
becomes
\begin{subequations}  \label{eq:AGS}   
\begin{align}  
\mcu_{11}  = {}&  - P_{34} \, (G_0 \, T \, G_0)^{-1}  -
P_{34} U_1 G_0 \, T \, G_0 \, \mcu_{11}  
+ U_2   G_0 \, T \, G_0 \, \mcu_{21}, \label{eq:U11}  \\
\label{eq:U21}
\mcu_{21} = {}& (1 - P_{34}) \, (G_0 \, T \, G_0)^{-1}  
+ (1 - P_{34}) U_1 G_0 \, T \, G_0 \, \mcu_{11},
\end{align}
\end{subequations}
where the subscripts 1 and 2 label the partitions of particles
(12,3)4 and (12)(34), i.e., the subsystems 3+1 and 2+2, respectively.
The operators in the above equations are
 the free resolvent
\begin{gather}\label{eq:g0}
  G_0 = (E+i0 -H_0)^{-1}
\end{gather}
 with the free Hamiltonian $H_0$,
 the two-nucleon pair (12) transition operator
\begin{gather}\label{eq:t}
T = V + V G_0 T
\end{gather}
with the two-nucleon potential $V$,
and the 3+1 or 2+2 subsystem transition operators
\begin{gather} \label{eq:AGSsub}
U_\alpha =  P_\alpha G_0^{-1} + P_\alpha T\, G_0 \, U_\alpha,
\end{gather}
with the respective permutation operators
$P_1 = P_{12}\, P_{23} + P_{13}\, P_{23}$ and $P_2 = P_{13}\, P_{24}$
being combinations of  permutation operators $P_{ij}$ of particles $i$ and $j$.

The elastic nucleon-trinucleon scattering amplitudes are calculated as  on-shell elements of $\mcu_{11}$, i.e.,
\begin{gather}\label{eq:ampl}
  \mct(\mbf{q}_f,\mbf{q}_i) = S_1 \langle \phi_1 \mbf{q}_f | \mcu_{11} | \phi_1 \mbf{q}_i \rangle,
\end{gather}
where the factor $S_1=3$ stems from the symmetrization,
$\mbf{q}_i$ ($\mbf{q}_f$) is the initial (final) relative nucleon-trinucleon momentum,
and $|\phi_1\rangle$ is the Faddeev amplitude of the three-nucleon bound state
$|\Phi_1 \rangle = (1+P_1) |\phi_1\rangle$. It is obtained solving the Faddeev equation
\begin{gather}\label{eq:phi}
  |\phi_1\rangle = G_0 T P_1 |\phi_1\rangle
\end{gather}
for the three-body  bound state with a proper normalization $ \langle \Phi_1 |\Phi_1 \rangle = 1$.

For the study of multiple scattering expansion  the AGS equations (\ref{eq:AGS}) are expanded into Neumann series
\begin{subequations}  \label{eq:AGSser}   
\begin{align}  
  \mcu_{11}^{(0)} = {}&  - P_{34} \, (G_0 \, T \, G_0)^{-1}, \\
  \mcu_{21}^{(0)} = {}& (1-P_{34})\, (G_0 \, T \, G_0)^{-1}, \\
  \mcu_{11}^{(n+1)} = {}& -P_{34} U_1 G_0 \, T \, G_0 \, \mcu_{11}^{(n)}  + U_2   G_0 \, T \, G_0 \, \mcu_{21}^{(n)}, \label{eq:U11n}  \\
\label{eq:U21n}
\mcu_{21}^{(n+1)} = {}& (1 - P_{34}) U_1 G_0 \, T \, G_0 \, \mcu_{11}^{(n)}
\end{align}
\end{subequations}
with $n=0,1,2,\ldots$.
The nucleon-trinucleon elastic scattering amplitude of the order $N$ in subsystem transition operators $U_\alpha$
is then obtained as
\begin{gather}\label{eq:TN}
  \mct^{(N)}(\mbf{q}_f,\mbf{q}_i) = S_1 \sum_{n=0}^N \langle \phi_1 \mbf{q}_f |\mcu_{11}^{(n)} | \phi_1 \mbf{q}_i \rangle.
\end{gather}
Explicitly, the contributions up to the second order are
\begin{subequations}  \label{eq:AGS012}   
  \begin{align} \label{eq:AGSu0}
    \mct^{(0)}(\mbf{q}_f,\mbf{q}_i) =  {}& -S_1 \, \langle \phi_1 \mbf{q}_f | P_{34} \, P_1 \, G_0^{-1}  | \phi_1 \mbf{q}_i \rangle, \\ \label{eq:AGSu1}
    \mct^{(1)}(\mbf{q}_f,\mbf{q}_i) =  {}& \mct^{(0)}(\mbf{q}_f,\mbf{q}_i) + S_1
    \langle \phi_1 \mbf{q}_f | [P_{34} U_1 P_{34} + U_2(1 - P_{34}) ] | \phi_1 \mbf{q}_i \rangle, \\ \nonumber
    \mct^{(2)}(\mbf{q}_f,\mbf{q}_i) =  {}& \mct^{(1)}(\mbf{q}_f,\mbf{q}_i) + S_1
    \langle \phi_1 \mbf{q}_f | \{ -P_{34} U_1 G_0 \, T \, G_0 [P_{34} U_1 P_{34} + U_2(1 - P_{34})] \\ {} &
    - U_2   G_0 \, T \, G_0 (1 - P_{34}) U_1 P_{34}   \} | \phi_1 \mbf{q}_i \rangle.
    \label{eq:AGSu2}
\end{align}
\end{subequations}
The negative power of $T$ in the $n=0$ term (\ref{eq:AGSu0}) was eliminated using Eq.~(\ref{eq:phi}).

The multiple scattering expansion of the amplitude (\ref{eq:ampl})
can be considered also from a different perspective,
namely, in terms of the two-particle transition operator $T$,
the first order contribution being of utmost importance for practical applications.
For this also the subsystem transition operators $U_\alpha$ in Eqs.~(\ref{eq:AGS}) and
(\ref{eq:AGS012}) are expanded in powers of $T$, i.e.,
\begin{gather} \label{eq:AGSsubt}
U_\alpha =  P_\alpha G_0^{-1} + P_\alpha T\, P_\alpha + \ldots .
\end{gather}
Remarkably, the expansion of the scattering
amplitude up to the first order in subsystem transition operators $U_\alpha$  (\ref{eq:AGSu1})
does not contain all contributions of the first order in $T$. Indeed,
the $P_\alpha G_0^{-1}$ term of (\ref{eq:AGSsubt}) inserted into  Eq.~(\ref{eq:AGSu2})
yields terms of  the first order in $T$.
Thus, the corresponding amplitude is obtained by using the two-term expansion  (\ref{eq:AGSsubt})
of  $U_\alpha$
in  Eq.~(\ref{eq:AGSu1}) but  one term only in  Eq.~(\ref{eq:AGSu2}), leading to
\begin{gather}  \label{eq:AGST}   
  \begin{split} 
    \mct^{(1,T)}(\mbf{q}_f,\mbf{q}_i) =  {}& S_1 \langle \phi_1 \mbf{q}_f |
    \big\{ (-P_{34} \, P_1 + P_2)(1 - P_{34}) \, G_0^{-1} \\  {}& 
+  [P_{34}  P_1 T\, P_1 P_{34} +   P_2 T\, P_2 (1 - P_{34}) ] \\ {} &
      -P_{34} P_1  \, T \, [P_{34} P_1 P_{34} + P_2(1 - P_{34})] 
    - P_2   \, T \,  (1 - P_{34}) P_1 P_{34}   \big\} | \phi_1 \mbf{q}_i \rangle.
\end{split}
\end{gather}
It was verified that terms containing $T$ add up to the properly antisymmetrized 
single-scattering amplitude
${}_A\langle \Phi_1 \mbf{q}_f | \sum_{j=1}^3 T_{j4} | \Phi_1 \mbf{q}_i \rangle_A$,
where the subscript $A$ denotes antisymmetrization and $T_{j4}$ is the two-particle
transition operator resulting from the interaction of the
 external nucleon 4 with any of the  internal nucleons $j$.

 All calculations, i.e., the exact solution of the AGS equations  (\ref{eq:AGS})
 and various approximations (\ref{eq:AGS012}) and (\ref{eq:AGST}), are performed
  in the momentum-space partial-wave representation, with the basis states \\
$ | k_x \, k_y \, k_z  \{l_z [(l_y \{[l_x (s_1 s_2)S_x]j_x \, s_3\}S_y ) J_y s_4 ] S_z\} \,\mathcal{JM} \rangle$ and \\
$|k_x \, k_y \, k_z  (l_z  \{ [l_x (s_1 s_2)S_x]j_x\, [l_y (s_3 s_4)S_y] j_y \} S_z) \mathcal{ J M} \rangle $
for 3+1 and 2+2 configurations, respectively. They are
characterized by three continuous variables
 $k_x , \, k_y$ and $k_z$ that are the magnitudes of the four-particle Jacobi momenta, 
the corresponding  orbital angular momenta
$l_x$, $l_y$, and $l_z$, and the spins $s_i$ of four nucleons, which are coupled to 
 the   total angular momentum  $\mathcal{J}$ and its projection  $\mathcal{M}$ via various 
 intermediate angular momenta  $S_x$, $S_y$, $S_z$, $j_x$,  $j_y$, and $J_y$.
The isospins of the four nucleons are coupled to the total isospin in the same order as the spins.
Due to the rotational and parity invariance the  AGS equations are independent of  $\mathcal{M}$
and do not couple states of different  $\mathcal{J}$ and total parity $\Pi = (-1)^{l_x+l_y+l_z}$.

Since the present work considers neutron-triton scattering at 
higher energy than in previous works, the states with considerably higher angular momenta
have to be included to achieve the convergence of results, i.e.,
$j_x,j_y \leq 5$, $l_x \leq 4$,  $l_y \leq 8$ (4 for 2+2 configurations), $l_z \leq 8$ (10 for nucleon-trinucleon states),
$J_y \leq 15/2$, and $\mathcal{J}\leq 11$.
The singularities in the integral equation kernel are treated using the 
complex-energy method with special weights \cite{deltuva:12c}. The values of the energy imaginary part
used for the extrapolation to the physical limit ${\rm Im}E \to +0$ range from 1.5 to 3.0 MeV.


\section{Results \label{sec:res}}

The study of the neutron-$\Hh$ scattering is performed  using  realistic high-precision
two-nucleon potential developed by Doleschall \cite{doleschall:04a,lazauskas:04a}.
At larger distances it has Yukawa tail but is nonlocal at short distances,
therefore often referred as the inside-nonlocal outside-Yukawa
(INOY04) potential. It is fitted simultaneously to the two-nucleon scattering data and
three-nucleon binding energy. This way some three-nucleon force contributions are
included implicitly, leading to the best description of four-nucleon observables
above 5 MeV  as compared to other realistic potentials
\cite{deltuva:12c,deltuva:14a,deltuva:15a,deltuva:13c,deltuva:15c}.
It was verified, however, that the qualitative features of the multiple-scattering series
and relative importance of various terms are independent of the choice of the realistic
two-nucleon potential.

The neutron-$\Hh$ multiple-scattering expansion is studied comparing results from the full
solution of the AGS equations (\ref{eq:AGS}) with approximations (\ref{eq:TN}) including
the subsystem transition operators $U_\alpha$ up to the first, second and third orders,
in the following abbreviated by U1, U2, and U3, respectively. The former two are given explicitly
in Eqs.~(\ref{eq:AGSu1}) and (\ref{eq:AGSu2}).
In addition are shown results based on  the approximation of the scattering
amplitude up to the first order in two-nucleon transition operator $T$, labeled by T1 and given
by Eq.~(\ref{eq:AGST}).  
The numerical results are openly available \cite{midas}.

\begin{figure}[!]
\includegraphics[scale=0.80]{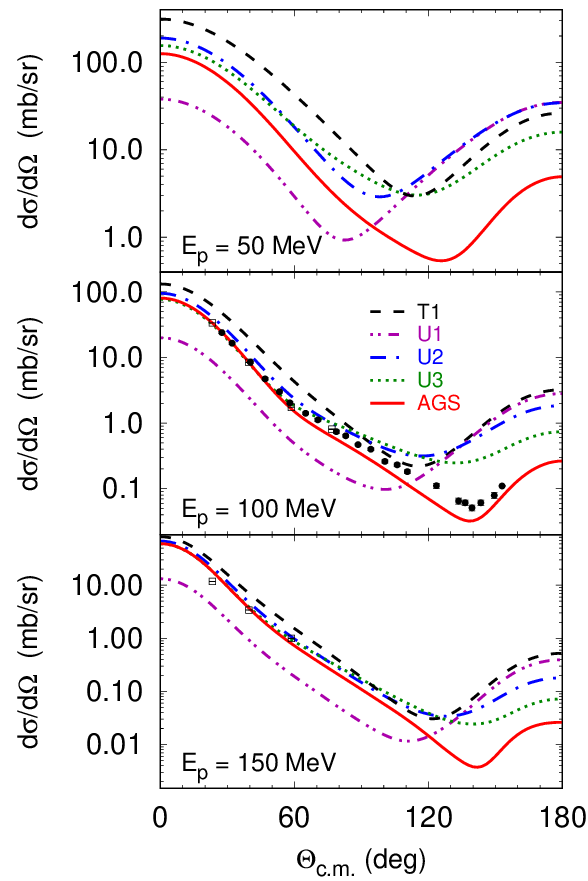}
\caption{\label{fig:dcs} (Color online)
  Differential cross section for neutron-$\Hh$ elastic scattering at 50, 100, and 150 MeV beam energy
  as a function of the c.m. scattering angle. Results of multiple scattering series truncated
  in several ways (see text) are compared with the full solution of AGS equations.
  The experimental proton-$\He$ data are from Refs.~\cite{ph100exp} (full circles) and 
  \cite{PhysRevC.32.1474} (open boxes).
}
\end{figure}

\begin{figure}[!]
\includegraphics[scale=0.80]{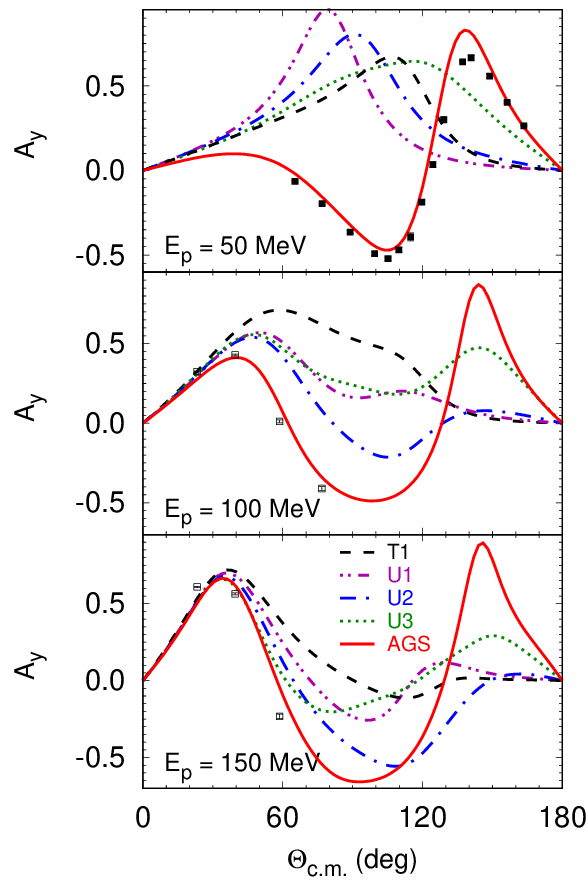}
\caption{\label{fig:ay} (Color online)
  Nucleon analyzing power for neutron-$\Hh$ elastic scattering at 50, 100, and 150 MeV beam energy
  as a function of the c.m. scattering angle. Results of multiple scattering series truncated
  in several ways (see text) are compared with the full solution of AGS equations.
  The experimental proton-$\He$ data are from Refs.~\cite{birchall:84a} (full boxes)
  and   \cite{PhysRevC.32.1474} (open boxes).
}
\end{figure}

The comparison is performed at three values of nucleon beam energies, i.e., 50, 100, and 150 MeV,
the calculated neutron-$\Hh$ elastic differential cross section $d\sigma/d\Omega$ 
and nucleon analyzing power $A_{\rm y}$  are shown in Figs.~\ref{fig:dcs} and \ref{fig:ay}
as functions of the scattering angle in the c.m. frame $\Theta_{\rm c.m.}$.
The approximation U1 fails heavily in reproducing the differential cross section at
all considered energies, and is the only one that underpredicts it at forward angles.
This is not surprising given the fact that U1 lacks some  contributions of the first order in $T$.
The other approximations U2, U3, and T1 tend to overpredict the $d\sigma/d\Omega$
results of the full AGS equations. At small angles the overprediction is sizable at 50 MeV
but decreases with increasing energy, such that at 100 and 150 MeV
the U3 results approach the exact ones at angles up to about $\Theta_{\rm c.m.} = 50^\circ$.
The lower-order approximations U2 and T1 at 150 MeV still lead to 10-20\% and 
30-40\% overestimation, respectively. At large angles, especially in the region of the minimum and
beyond it at backward angles, all approximations significantly overpredict the 
differential cross section from exact calculations. Thus, in this regime the
scattering amplitude  appears to be a result of a delicate interplay of many multiple scattering terms
up to a high order, the truncation of the series at a low order is inadequate and the
exact solution of four-body scattering equations is needed for reliable predictions.
Notably, the exchange term alone contained in Eq.~(\ref{eq:AGSu0}) roughly reproduces the 
cross section at very backward angles.

Qualitatively similar conclusions can be drawn from the comparison of the
nucleon analyzing power  in Fig.~ \ref{fig:ay}, and are typical for most spin observables.
At the lowest energy of 50 MeV all approximations are unable to reproduce even the shape of $A_{\rm y}$,
but with increasing energy are getting closer to the exact results at small angles
up to the first local maximum and slightly beyond it.
The failure at intermediate and backward angles remains at all energies.

Where available the predictions are compared with the experimental data for the charge-symmetric
proton-$\He$ reaction \cite{ph100exp,PhysRevC.32.1474,birchall:84a}.
The  overall agreement is quite reasonable, more significant underprediction of the
differential cross section is seen at 100 MeV around the minimum region at large angles,
consistently with previous findings below 40 MeV  \cite{deltuva:13c}.

\begin{figure}[!]
\includegraphics[scale=0.80]{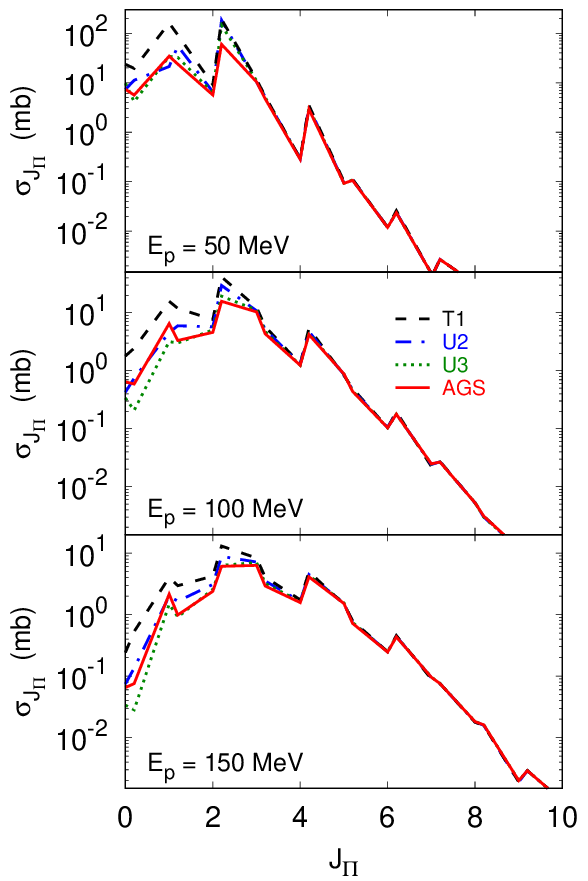}
\caption{\label{fig:sig} (Color online)
  Contributions to the neutron-$\Hh$ total elastic cross section at 50, 100, and 150 MeV beam energy
  from states with different total  angular momentum and parity.
 Results of multiple scattering series truncated
  in several ways (see text) are compared with the full solution of AGS equations.
}
\end{figure}

In order to get insight into the multiple-scattering expansion for each partial wave,
the  neutron-$\Hh$ total elastic cross section $\sigma_e$
at 50, 100, and 150 MeV is decomposed into the 
contributions  $\sigma_{J_\Pi}$ with fixed total angular momentum $\mathcal{J}$ and parity $\Pi$.
The positive and negative parity in Fig.~\ref{fig:sig} 
is distinguished by introducing the parameter
$J_\Pi = \mathcal{J} + (1-\Pi)/10$, that is, negative parity results are shown  at the
$x$-axis value shifted by 0.2. 
At low values of  $\mathcal{J}$ (or $J_\Pi$) with largest $\sigma_{J_\Pi}$
none of the considered
approximations U2, U3, and T1 is accurate, with T1 showing the largest deviations.
However, for larger $\mathcal{J}$ all of them are approaching the results from
the exact solution of AGS equations.
U1 approximation fails heavily for all partial waves and it is not shown.
Noteworthy,  there is no evident improvement with increasing energy 
in the convergence of the truncated multiple scattering series U2, U3, and T1
toward the exact results. This may seem to be inconsistent with the findings for
scattering observables in  Figs.~\ref{fig:dcs} and \ref{fig:ay}. 
However, the relative importance of different partial waves changes with energy,
namely, the decrease of $\sigma_{J_\Pi}$ with increasing $\mathcal{J} > 4$
is considerably slower at higher energy as Fig.~\ref{fig:sig} shows.
This is simply the manifestation of the well known feature that the number of the
contributing partial waves increases with energy.
The increasing importance of large $\mathcal{J}$ amplitudes together with
a good accuracy of  U2, U3, and T1 approximations at large $\mathcal{J}$ leads
to an improved agreement with exact results  for scattering observables
at higher energies, seen in Figs.~\ref{fig:dcs} and \ref{fig:ay}.

\begin{figure}[!]
\includegraphics[scale=0.80]{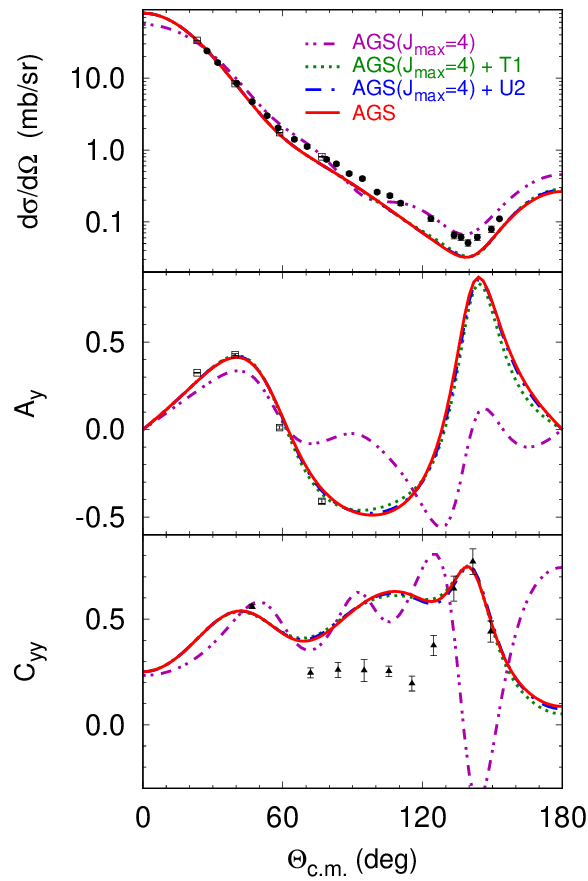}
\caption{\label{fig:j} (Color online)
  Differential cross section, nucleon analyzing power $A_{\rm y}$, and
  beam-target spin correlation coefficient $C_{\rm yy}$
  for neutron-$\Hh$ elastic scattering at  100 MeV beam energy
  as functions of the c.m. scattering angle.
  All results use exact scattering amplitudes for  $\mathcal{J} \leq J_{\rm max} = 4$ partial waves,
  but differ in the treatment of higher waves up to  $\mathcal{J} = 11$:
  neglected (double-dotted-dashed curves),
  taken in T1 (dotted curves)  or U2 (dashed-dotted curves) approximations, and
  calculated exactly (solid curves).
The experimental proton-$\He$ data are from Ref.~\cite{ph100exp,PhysRevC.32.1474,PhysRevC.106.054002}. 
}
\end{figure}

The above observation suggests a possible practical ``hybrid'' approach for the calculation of
scattering amplitudes. Namely, for low $\mathcal{J} \leq J_{\rm max}$ values the AGS equations are
solved exactly, but for $\mathcal{J} > J_{\rm max}$ a low-order approximation such as T1 or U2
is used. Figure \ref{fig:j} demonstrates the accuracy of such an approach
for the differential cross section, nucleon analyzing power, and
   neutron-$\Hh$ spin correlation coefficient $C_{\rm yy}$
   at 100 MeV nucleon beam energy. The results including only partial waves with
   $\mathcal{J} \leq J_{\rm max} = 4$ show several oscillations and are obviously
   inadequate. However, the ``hybrid'' amplitudes with added  higher $\mathcal{J}$ components in T1 or U2
   approximations lead to a good reproduction of exactly calculated observables,
   even in the region of the differential cross section minimum.

\section{Conclusions \label{concl}}

The elastic neutron-$\Hh$ scattering at intermediate energies was studied
using rigorous integral equations for four-particle transition operators.
They were expanded into multiple-scattering series. Truncation of the series in different ways
led to approximate scattering amplitudes whose calculation is simpler and less demanding
as compared to the exact solution of scattering equations.
 All calculations were performed in the momentum-space
partial-wave representation.

The studied approximations are the expansions of the  neutron-$\Hh$ scattering
amplitudes in terms of 3+1 and 2+2 subsystem transition operators up to the first, second, and third order,
and in terms of the two-nucleon transition operator up to the first order,
labeled U1, U2, U3, and T1, respectively.
The validity of those approximations was investigated by comparing their
results for the differential cross section and nucleon analyzing power
with the reference ones based on the exact solution.
At lower energy of 50 MeV all the approximations fail, but, except for U1, 
at higher energies around 150 MeV provide a rough reproduction of exact results at small scattering
angles up to about 50 deg. Noteworthy, the quality of agreement correlates with the sophistication
of the approximation, i.e., U3 is better than U2 which is better than T1.
In the large-angle region with small differential cross section
all approximations fail heavily, indicating that the
scattering amplitude results from a delicate interplay of many multiple scattering terms
up to a high order and 
exact solution of four-body scattering equations is needed for reliable predictions.
The study of partial-wave total elastic cross sections further suggests that,
independently of the energy, this
interplay of multiple scattering terms takes place in lower partial waves only, typically,
with four-nucleon total angular momentum $\mathcal{J} \leq J_{\rm max} = 4$,
 while for higher $\mathcal{J}$ the 
 U2, U3, and T1 approximations appear to be quite accurate. This feature,
 together with the increasing relative importance of 
 higher partial waves with increasing energy, explains why
 the reproduction of reference results by 
 U2, U3, and T1 approximations  improves with rising energy.
 Furthermore, based on this feature a ``hybrid'' approach was proposed,
 that combines exact amplitudes in lower partial waves and T1 or U2 approximations in higher
 partial waves. Such results are very accurate but the computations are less time consuming,
 rendering this approach useful for practical calculations.

 Regarding the  approximation of the first order in the two-nucleon transition operator,
  it reasonably reproduces the shape of the differential cross section
 at small angles, but overestimates its magnitude. Thus, its application to scattering processes
 has to be considered with some care. The microscopic optical potential built from this
 approach  still has to be iterated in the Lippmann-Schwinger equation
 and may be more successful. On the other hand, the present
 work indicates that the potential should be perturbative in higher partial waves, with accurate results
 in Born approximation. The development of such a potential for the neutron-$\Hh$ scattering
and its benchmark is in progress \cite{vorabbi:pc}.

\vspace{1mm}
\begin{acknowledgments}
  Author thanks R. Crespo, Ch. Elster, N. Timofeyuk, and M. Vorabbi for discussions.
This work has received funding from the 
Research Council of Lithuania (LMTLT) under Contract No.~S-MIP-22-72.
Part of the computations were performed using the infrastructure of
the Lithuanian Particle Physics Consortium.
\end{acknowledgments}


\begin{thebibliography}{10}

\bibitem{nollett:07a}
K.~M. Nollett, S.~C. Pieper, R.~B. Wiringa, J. Carlson, and G.~M. Hale, Phys.
  Rev. Lett. {\bf 99},  022502  (2007).

\bibitem{quaglioni:08a}
S. Quaglioni and P. Navratil, Phys. Rev. Lett. {\bf 101},  092501  (2008).

\bibitem{navratil:12a}
P. Navr\'atil and S. Quaglioni, Phys. Rev. Lett. {\bf 108},  042503  (2012).

\bibitem{hupin:19a}
G. Hupin, S. Quaglioni, and P. Navr\'atil, Nature Communications {\bf 10},  351
   (2019).

\bibitem{lazauskas:18a}
R. Lazauskas, Phys. Rev. C {\bf 97},  044002  (2018).

\bibitem{deltuva:12c}
A. Deltuva and A.~C. Fonseca, Phys.~Rev.~C {\bf 86},  011001(R)  (2012).

\bibitem{deltuva:14a}
A. Deltuva and A.~C. Fonseca, Phys.~Rev.~Lett. {\bf 113},  102502  (2014).

\bibitem{deltuva:15a}
A. Deltuva and A.~C. Fonseca, Phys. Lett. B {\bf 742},  285  (2015).

\bibitem{lazauskas:15a}
R. Lazauskas, Phys. Rev. C {\bf 91},  041001(R)  (2015).

\bibitem{CRESPO200626}
R. Crespo, A. Moro, and I. Thompson, Nuclear Physics A {\bf 771},  26  (2006).

\bibitem{PhysRevC.74.044616}
R. Crespo, I.~J. Thompson, and A.~M. Moro, Phys. Rev. C {\bf 74},  044616
  (2006).

\bibitem{microOP:23}
C. Hebborn {\it et~al.}, Journal of Physics G: Nuclear and Particle Physics
  {\bf 50},  060501  (2023).

\bibitem{gennari:18a}
M. Gennari, M. Vorabbi, A. Calci, and P. Navr\'atil, Phys. Rev. C {\bf 97},
  034619  (2018).

\bibitem{vorabbi:22a}
M. Vorabbi, M. Gennari, P. Finelli, C. Giusti, P. Navr\'atil, and R. Machleidt,
  Phys. Rev. C {\bf 105},  014621  (2022).

\bibitem{furumoto:mgop}
T. Furumoto, K. Tsubakihara, S. Ebata, and W. Horiuchi, Phys. Rev. C {\bf 99},
  034605  (2019).

\bibitem{PhysRevC.102.034606}
M. Burrows, R.~B. Baker, C. Elster, S.~P. Weppner, K.~D. Launey, P. Maris, and
  G. Popa, Phys. Rev. C {\bf 102},  034606  (2020).

\bibitem{PhysRevC.109.034613}
M. Vorabbi, C. Barbieri, V. Som\`a, P. Finelli, and C. Giusti, Phys. Rev. C
  {\bf 109},  034613  (2024).

\bibitem{PhysRevC.110.034605}
R.~B. Baker, C. Elster, T. Dytrych, and K.~D. Launey, Phys. Rev. C {\bf 110},
  034605  (2024).

\bibitem{deltuva:13c}
A. Deltuva and A.~C. Fonseca, Phys.~Rev.~C {\bf 87},  054002  (2013).

\bibitem{deltuva:15c}
A. Deltuva and A.~C. Fonseca, Phys.~Rev.~C {\bf 91},  034001  (2015).

\bibitem{PhysRevC.106.054002}
A. Watanabe {\it et~al.}, Phys. Rev. C {\bf 106},  054002  (2022).

\bibitem{vorabbi:pc}
  M. Vorabbi, private communication (2024).

\bibitem{grassberger:67}
P. Grassberger and W. Sandhas, Nucl. Phys. {\bf B2},  181  (1967); E. O. Alt,
  P. Grassberger, and W. Sandhas, JINR report No. E4-6688 (1972).

\bibitem{yakubovsky:67}
O.~A. Yakubovsky, Yad. Fiz. {\bf 5},  1312  (1967) [Sov. J. Nucl. Phys. {\bf
  5}, 937 (1967)].

\bibitem{viviani:11a}
M. Viviani, A. Deltuva, R. Lazauskas, J. Carbonell, A.~C. Fonseca, A. Kievsky,
  L.~E. Marcucci, and S. Rosati, Phys.~Rev.~C {\bf 84},  054010  (2011).

\bibitem{viviani:17a}
M. Viviani, A. Deltuva, R. Lazauskas, A.~C. Fonseca, A. Kievsky, and L.~E.
  Marcucci, Phys. Rev. C {\bf 95},  034003  (2017).

\bibitem{viviani:01a}
M. Viviani, A. Kievsky, S. Rosati, E.~A. George, and L.~D. Knutson, Phys. Rev.
  Lett. {\bf 86},  3739  (2001).

\bibitem{kievsky:08a}
A. Kievsky, S. Rosati, M. Viviani, L.~E. Marcucci, and L. Girlanda, J. Phys. G
  {\bf 35},  063101  (2008).

\bibitem{lazauskas:04a}
R. Lazauskas and J. Carbonell, Phys. Rev. C {\bf 70},  044002  (2004).

\bibitem{lazauskas:09a}
R. Lazauskas, Phys. Rev. C {\bf 79},  054007  (2009).

\bibitem{deltuva:07a}
A. Deltuva and A.~C. Fonseca, Phys.~Rev.~C {\bf 75},  014005  (2007).

\bibitem{doleschall:04a}
P. Doleschall, Phys.~Rev.~C {\bf 69},  054001  (2004).

\bibitem{midas}
https://midas.lt/public-app.html\#/research/dataObjects?resourceId=258757\&uuid=6f1abf88-e451-49d8-85ae-9a6612372f31\&lang=en

\bibitem{ph100exp}
N.~P. Goldstein, A. Held, and D.~G. Stairs, Canadian Journal of Physics {\bf
  48},  2629  (1970).

\bibitem{PhysRevC.32.1474}
J.~S. Wesick, P.~G. Roos, N.~S. Chant, C.~C. Chang, A. Nadasen, L. Rees, N.~R.
  Yoder, A.~A. Cowley, S.~J. Mills, and W.~W. Jacobs, Phys. Rev. C {\bf 32},
  1474  (1985).

\bibitem{birchall:84a}
J. Birchall, W.~T.~H. van Oers, J.~W. Watson, H.~E. Conzett, R.~M. Larimer, B.
  Leemann, E.~J. Stephenson, P. von Rossen, and R.~E. Brown, Phys. Rev. C {\bf
  29},  2009  (1984).

\end{thebibliography}

 \end{document}